\documentclass[aps,prb,floatfix,twocolumn,showpacs,superscriptaddress,groupedaddress]{revtex4-2}
\usepackage{graphicx}
\usepackage{amsmath, amssymb}
\usepackage{xcolor}
\usepackage{amsfonts}
\usepackage{hhline}
\usepackage{braket}
\usepackage{bm}
\usepackage{comment}
\usepackage{mathrsfs}
\usepackage{hyperref}
\usepackage[normalem]{ulem} 
\hypersetup{colorlinks=true,breaklinks,linkcolor=blue,urlcolor=blue,citecolor=blue}

\begin{document}
\title{
Semi-automated estimation of hydrogenic initial states for localized Wannier functions
}
\author{Tatsuki Oikawa$^{1*}$}
\author{Kota Ido$^{2*}$}
\author{Takahiro Misawa$^2$}
\author{Takashi Koretsune$^1$} 
\author{Kazuyoshi Yoshimi$^2$}
\affiliation{
$^1$Department of Physics, Tohoku University, Sendai 980-8578, Japan \\
$^2$Institute for Solid State Physics, University of Tokyo, Kashiwa, Chiba 277-8581, Japan \\
$^*$ These authors contributed equally to this work.
}
\begin{abstract}
We present a semi-automated method for obtaining an initial estimate of Wannier functions, 
designed to facilitate the construction of Wannier functions for describing low-energy effective models of 
solids, particularly those relevant to strongly correlated electron systems. 
Our approach automatically determines the hydrogenic projections
orbitals and the center of the Wannier functions 
from information on Bloch wavefunctions at the $\Gamma$ point.
This method is integrated into cif2qewan, 
enabling seamless generation of input files for Quantum ESPRESSO and Wannier90. 
We validate our method through applications to both inorganic and organic compounds, such as Si, SrVO$_3$, FeSe, Na$_8$Al$_6$Si$_6$O$_{24}$, and (TMTTF)$_2$PF$_6$.
The obtained results demonstrate that our semi-automated projections give a good initial estimate of the Wannier functions. 
We also show the comparisons with other methods for estimating the initial states of the Wannier functions, such as the Selected Columns of the Density Matrix (SCDM).
Our methodology shows an efficient way to construct Wannier functions, 
paving the way for high-throughput calculations in the study of complex materials.
\end{abstract}

\maketitle
\section{Introduction}
Wannierization---constructing localized Wannier functions describing low-energy degrees of freedom in materials---is a crucial step in understanding electronic structures of materials~\cite{Marzari_RMP2012,Marrazzo_RMP2024}. The finding of an efficient method for constructing the maximally localized Wannier functions~\cite{Marzari_PRB1997,Souza_PRB2001} has induced an essential advancement in electronic structure calculations, such as accurate interpolations of band structures~\cite{Yates_PRB2007}, calculations of anomalous Hall conductivities~\cite{Wang_PRB2006}, and the derivation of low-energy effective Hamiltonians of strongly correlated systems~\cite{Aryasetiawan_PRB2004,Imada_JPSJ2010}. The Wannierization not only improves computational efficiency but also enhances physical interpretability by providing an intuitive real-space representation of electronic states.

The development of software packages for the construction and use of Wannier functions is rapidly progressing. Wannier90~\cite{Mostofi_CPC2008,Mostofi_CPC2014,Pizzi_2020,wan90_HP} is now a common tool for constructing Wannier functions, and several software packages for $ab$ $initio$ calculations, such as Quantum ESPRESSO~\cite{QE,QE_HP}, VASP~\cite{VASP_1,VASP_2}, WIEN2k~\cite{WIEN2K}, ABINIT~\cite{ABINIT}, and OpenMX~\cite{Ozaki_PRB2003,Ozaki_PRB2004,OpenMX_HP} have an interface to Wannier90. Many software packages have also been developed to calculate various physical quantities using Wannier functions~\cite{Wannier_Eco}. 

In this way, Wannier functions are widely used in various fields of condensed matter physics.
However, constructing Wannier functions often requires careful attention, particularly in choosing appropriate initial conditions.
A standard approach is to use projections onto hydrogen-like atomic orbitals (hydrogenic projection)~\cite{Marzari_PRB1997, Souza_PRB2001}.
By properly defining initial orbitals based on the constituent elements, this method has been successfully employed in high-throughput calculations across a wide range of materials~\cite{cif2qewan_hp,Sakai2020,Kurita2020,Garrity2021}.
More recently, projectability-based disentanglement schemes have also been developed~\cite{Qiao_PDWF}.
Furthermore, the Manifold-Remixed Wannier Function (MRWF) method~\cite{Qiao_MRWF} has enabled the construction of Wannier functions for target bands by automatically selecting suitable combinations of orbitals from the original Wannier basis.
However, as long as one starts from hydrogenic projections, Wannier functions are inherently limited due to their reliance on predefined atomic orbitals.
Consequently, such approaches are generally not suitable for capturing interstitial states or molecular orbitals.

To overcome these limitations, wavefunction-based approaches have been proposed, in which the initial guesses for Wannier functions are generated directly from the Bloch wavefunctions.
One prominent example is the Selected Columns of the Density Matrix (SCDM) method~\cite{Damle_MMS2018, Damle_JCC2015}, which identifies an optimal subspace using a quasi-density matrix.
This approach enables the automatic and efficient construction of nontrivial Wannier functions and has been successfully applied to both isolated and entangled bands in high-throughput workflows~\cite{vitale2020automated}.
However, a potential drawback is that the resulting Wannier functions may significantly deviate from atomic-like orbitals, making physical interpretation difficult.

In this work, we propose a semi-automated method for constructing initial states of Wannier functions that lies between these two regimes, that is, automatically predicting hydrogenic orbitals directly from the Bloch wavefunctions.
In this method, hydrogenic projections are determined based on information of Bloch functions at the $\Gamma$ point. The projection positions are set at the center of the charge density for each cluster. This procedure allows both the projected atomic orbitals and their centers to be determined automatically. An advantage of this method is that it can construct Wannier functions that are relatively close to atomic orbitals, which makes an intuitive understanding of electronic structures easier. 
Generally, the selection of bands for Wannierization needs to be done manually. However, in cases where bands are well-isolated, such as in organic compounds, this selection can also be automated. We will demonstrate that the method works well for constructing the Wannier functions for several prototypical compounds, such as a semiconductor Si, a correlated metal SrVO$_{3}$, an iron-based superconductor FeSe, a sodalite Na$_8$Al$_6$Si$_6$O$_{24}$, and an organic compound (TMTTF)$_2$PF$_{6}$.
For easily performing this method, we also developed open-source software, SEAP(Semi-automated Estimator for Atomic Projections of Wannier functions)~\cite{seap_hp}.
This software can be combined with cif2qewan~\cite{cif2qewan_hp}, a tool that generates input files for Quantum ESPRESSO and Wannier90 from CIF (Crystallographic Information File). 

This paper is organized as follows:
In Section \ref{sec:method}, we explain our semi-automated method for constructing initial states of Wannier functions. 
In Section \ref{sec:applications},
we demonstrate how the method works for constructing the Wannier functions in Si, SrVO$_3$, FeSe, Na$_8$Al$_6$Si$_6$O$_{24}$, and
(TMTTF)$_{2}$PF$_{6}$. In particular, we show that
the method can estimate the Wannier centers even for the complicated compounds (TMTTF)$_{2}$PF$_{6}$.
Section \ref{sec:summary} is devoted to the summary and discussion.

\section{Method}
\begin{figure*}[t]
  \centering
  \resizebox*{16cm}{!}{\includegraphics{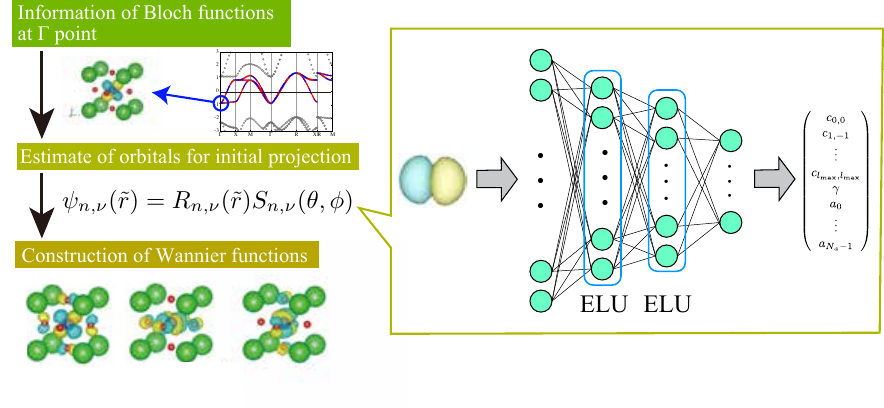}}
  \caption{Schematic workflow of the proposed semi-automated method for constructing Wannier functions.
  The details are given in Sec. \ref{sec:method} of the main text.
  } \label{fig:workflow}
\end{figure*}

In this section, we explain our semi-automated method for constructing initial states of Wannier functions. The schematic workflow is shown in Fig.~\ref{fig:workflow}. The method is composed of three steps: extracting information of Bloch functions at the $\Gamma$ point, estimating orbitals for hydrogenic projections using neural networks, and determining the initial projections from the estimated results. Each step is explained in detail below.
\label{sec:method}

\subsection{Post-processing procedure for wavefunction data} \label{sec:method-postprocess}
First, we overview the post-processing procedure applied to Bloch functions at the $\Gamma$ point. 
We extract information on the lattice vectors, atomic coordinates, and grid-based wavefunction data
from the information of the Bloch functions at the $\Gamma$ point.
Then, a clustering algorithm is applied to classify atoms into molecular clusters 
based on a bond-length threshold criterion.  
Finally, the wavefunction data are interpolated, 
resulting in a refined spatial distribution of the electronic density. 
These interpolated wavefunctions are used to estimate the projections.
Each of these processes is described in detail below.

To begin with, we extract information about Bloch functions at the $\Gamma$ point for target bands. 
For norm-conserving pseudopotentials~\cite{ppstype_NC}, from Bloch functions at
$\Gamma$ point $\psi_{n}(\bm{r})$, we can obtain the following charge distribution
\begin{align}
    |\psi_{n}(\bm{r})|^{2}\text{sign}(\psi_{n}(\bm{r})),
\end{align}
where $n$ {represents} the band index.
Based on the charge distribution, we automate the hydrogenic projections and the determination of the orbital centers.
We note that the above expression holds only for norm-conserving pseudopotentials.
For the ultrasoft pseudopotentials, strictly speaking, 
the Bloch functions at the $\Gamma$ point can differ from the charge distributions.
However, it is plausible that the Bloch functions at the $\Gamma$ point have similar quantities even for ultrasoft pseudopotentials.

To systematically determine the projection candidates' centers, we perform a clustering procedure that groups atoms 
and wavefunction data points based on their spatial proximity. 
The clustering process consists of the following three steps: 
(1) atomic clustering based on interatomic distances, 
(2) assigning the wavefunction charge density to the identified clusters, 
and (3) computing the center of each cluster using a charge density-weighted average.

Given a set of atomic coordinates within a unit cell, 
we classify atoms into clusters according to their spatial proximity. 
Specifically, two atoms are considered to belong to the same cluster 
if their interatomic distance is less than a predefined threshold $d_{\text{th}}$. 
Two atoms at positions $\boldsymbol{r}_i$ and $\boldsymbol{r}_j$
are assigned to the same cluster $\nu$ if the following condition is satisfied.
\begin{equation}
    |\boldsymbol{r}_i - \boldsymbol{r}_j| < d_{\text{th}}.
\end{equation}
This criterion is applied iteratively to form larger clusters 
by grouping atoms that are transitively connected. 
The threshold $d_{\text{th}}$ is chosen based on the typical bond lengths in the system, 
ensuring that chemically or structurally 
relevant atomic groupings are preserved.
In actual calculations presented in the next section, we set $d_{\text{th}}$ to be 1.5 {\AA} for inorganic compounds and 2.0 {\AA} for organic compounds, unless otherwise noted. 

For most compounds, Bloch wavefunctions are distributed near the atoms.
However, in some compounds such as electrides and nanoporous materials, these are located near interstitial sites.
For such cases, we should add interstitial sites to a set of atoms before the clustering procedure mentioned above.
To add interstitial sites, we use the Voronoi tessellation method. 
In this paper, we set the threshold of the distance between atoms and interstitial sites $d_{{\rm th}}^{{\rm ai}}$ to $2.0{\rm \AA}$.

Once atomic clusters are identified, 
we assign the charge distribution to these clusters. 
Given a discrete wavefunction grid, each grid point $\boldsymbol{r}_i$ 
is assigned to the nearest cluster based on the Euclidean distance 
between the grid point and the nearest atomic position within a cluster. 
This step ensures that the electron density is 
spatially assigned to the most relevant atomic environment.

To quantitatively characterize the spatial distribution of charge density within each cluster, we compute the charge-density-weighted center of each cluster. 
For a given cluster $\nu$, the center position $\widetilde{\boldsymbol{r}}^{(n,\nu)}$  is defined as
\begin{equation}\label{cdwc}
    \widetilde{\boldsymbol{r}}^{(n,\nu)} = \frac{\sum\limits_{i \in \nu} |\psi_{n}(\boldsymbol{r}_i)|^2 \boldsymbol{r}_i}{\sum\limits_{i \in\nu} |\psi_{n}(\boldsymbol{r}_i)|^2}.
\end{equation}
Here, $|\psi_n(\boldsymbol{r}_i)|^2$ represents the squared norm of 
the wavefunction at grid point $\boldsymbol{r}_{i}$, 
which corresponds to the local charge density contribution at that point. 
The summation runs over all the data points belonging to the cluster $\nu$. 
This weighted center formulation ensures that regions with high electrondensity 
contribute more significantly to the cluster's effective position. 
At this stage, we can safely discard clusters where sum of the charge density, $\rho_\nu = \sum_{\boldsymbol{r}, i \in\nu} |\psi_{n}(\boldsymbol{r}_i)|^2$, is much smaller than that of other clusters. 
In actual calculations, we only keep clusters where $\rho_\nu$ is larger than 10\% of the total charge in the whole unit cell. 

Furthermore, high-resolution three-dimensional data is generated by interpolating the wavefunction values within a cubic region around the center of the electron density. This process converts raw wavefunction data into a grid-based volumetric representation suitable for visualization using software such as VESTA\cite{vesta}. 
In this paper, the length of the cubic region $l_{\rm c}$ is set to the same value as the bond length threshold $d_{\rm th}$ for inorganic compounds, and $2d_{\rm th}$ for organic compounds. For evaluating quantities in Eq.~\eqref{cdwc}, we need to associate each spatial point with its corresponding molecular cluster. 
To this end, we apply the $k$-dimensional tree ($k$d-tree) algorithm\cite{bentley1975multidimensional,maneewongvatana1999s} to assign each point on the fine interpolation mesh to the cluster of its nearest atom, enabling accurate cluster-resolved analyses.
The computed spatial distributions and electron density centers are saved in the xsf format, 
while numerical data is stored in the csv and npy formats 
for further computational analysis.

The final output consists of the following three datasets: (1) electron density centers of molecular clusters; 
(2) a three-dimensional wavefunction grid dataset suitable for machine learning applications; 
and (3) band-resolved electron density distributions. 
The dataset in (2) is used for the projection of electron states, 
as described in the following sections.

\subsection{Estimation of orbitals for projections}
To determine the orbitals for the projections, we assume the Bloch functions
can be expressed as a product of the radial functions $R_{n,\nu}(\tilde{r})$ and the real spherical harmonics $S_{n,\nu}(\theta,\phi)$: 
\begin{align}
        \label{eq:psi}
        \psi_{n,\nu}(\tilde{\bm{r}})=R_{n,\nu}(\tilde{r})S_{n,\nu}(\theta, \phi),
\end{align}
where $\tilde{\bm{r}}=\bm{r}-\tilde{\bm{r}}_{\nu}^{n}$, 
$\tilde{r}=|\tilde{\bm{r}}|$.
Here, $S_{n,\nu}(\theta, \phi)$ and $R_{n,\nu}(\tilde{r})$ are defined as
\begin{align}
        \label{eq:S}
        &S_{n,\nu}(\theta, \phi) = \sum^{l_{\text{max}}}_{l=0}\sum^{l}_{m=-l}c^{(n,\nu)}_{lm}S_{lm}(\theta,\phi),\\
        \label{eq:R}
        &R_{n,\nu}(\tilde{r}) = e^{-\gamma^{(n,\nu)}\tilde{r}}\sum^{N_{a}-1}_{p=0} a^{(n,\nu)}_{p} \tilde{r}^{p},
\end{align}
where $n$ is the band index, $\nu$ is the cluster index, 
$l$ is the azimuthal quantum number, and 
$m$ is the magnetic quantum number. 
The cutoff parameter $N_a$ specifies the number of polynomial basis functions used in the radial expansion.
In practice, $N_a$ determines the flexibility of the radial function:
larger $N_a$ allows for a more accurate representation at the cost of higher computational complexity.
The coefficients $a^{(n,\nu)}_{p}$ are determined so as to satisfy the normalization condition, and the factor $\gamma^{(n,\nu)}$ controls the exponential decay of the radial tail.
For simplicity, hereafter we omit the cluster index $\nu$ of the coefficients $c^{(n,\nu)}_{lm}$, $\gamma^{(n,\nu)}$, and $a^{(n,\nu)}_{p}$ unless otherwise noted.
One can take the orbitals with large expansion coefficients $c^{(n)}_{lm}$ 
as the initial guess for the Wannier functions.
However, it is difficult to optimize the radii and coefficients using a simple linear regression.
In particular, we find that the determination of the radius $R_{n}(\tilde{r})$ is difficult.

To efficiently estimate coefficients in the radial function and the spherical harmonic basis, we construct and train a neural network (NN)-based encoder from wavefunction data. 
The encoder is designed as a fully connected network (FCN) consisting of 
three linear transformation layers. 
Exponential Linear Unit (ELU) was adopted as the activation function to mitigate vanishing gradient problems while maintaining nonlinearity\cite{clevert2016fastaccuratedeepnetwork}. The input to the model is wavefunction data discretized on a three-dimensional grid obtained from first-principles calculations, 
while the output consists of 16 expansion coefficients corresponding to 
spherical harmonics up to an angular momentum of $l_{\rm max} = 3$, 
together with the radial parameters $\gamma^{(n)}$ and $\{a^{(n)}_{p}\}_{p=0}^{N_a-1}$,
where the number of radial basis functions is fixed at $N_a = 4$:
\begin{align}
  \label{eq:params}
  \left(
  c^{(n)}_{0,0}, c^{(n)}_{1,-1}, \ldots, c^{(n)}_{l_{\rm max}, l_{\rm max}},
  \ \gamma^{(n)},\ a^{(n)}_{0}, a^{(n)}_{1}, a^{(n)}_{2}, a^{(n)}_{3}
  \right).
\end{align}
The loss function is defined as the mean squared error (MSE) between the predicted and reference wavefunction data 
discretized on the three-dimensional grid, instead of between the expansion coefficients:
\begin{align}
        \label{eq:loss}
        L = \frac{1}{N^3}\sum_{i}
        \left[\psi^{\text{out}}_{n,\nu}(\bm{r}_{i})
        - \psi^{\text{in}}_{n,\nu}(\bm{r}_{i})\right]^{2},
\end{align}
where $N$ means the number of grids per side, $\psi^{\text{in}}_{n,\nu}$ denotes the input wavefunction obtained from first-principles calculations,
and $\psi^{\text{out}}_{n,\nu}$ is the reconstructed wavefunction computed from the predicted parameters
using Eqs.~\eqref{eq:psi}, \eqref{eq:S}, and \eqref{eq:R}.
This wavefunction-based loss was adopted because it provided higher reconstruction accuracy and more stable convergence than directly minimizing the coefficient error.
We found that using the wavefunction-based loss yielded higher reconstruction accuracy and more stable convergence than directly comparing the coefficients.
For training, we generated a dataset 
of $20000$ wavefunction samples, 
where each wavefunction was represented on a 
$N\times N\times N$ ($N=32$) grid within a box with sides of $10.0$ \AA. 
The dataset was split into training (81\%), validation (9\%), and test (10\%) sets, 
and training was performed using mini-batches (the batch size is 128). 
By minimizing the loss function defined in Eq.~(\ref{eq:loss}), we optimize the model parameters using the Adam optimizer~\cite{optimizer_Adam} 
and evaluate the derivative of the loss function using forward differences.
The model was trained for 300 epochs until convergence.
After training, the model was evaluated on the validation and test datasets. 
This encoder serves as a pre-trained neural network model 
that enables efficient and automated expansion of wavefunctions 
into a spherical harmonic basis.

Using the trained neural network, we estimate the expansion coefficients from discretized wavefunction data. Given an input wave function represented on a three-dimensional grid, the neural network extracts spatial features through 
its hidden layers and maps them to the corresponding coefficients in Eq. \eqref{eq:params}: the network generates the radial function parameters $\gamma^{(n)}$, 
the polynomial coefficients $a_{p}^{(n)}$, and
the real-valued spherical harmonic expansion coefficients $c_{lm}^{(n)}$.
These coefficients are obtained through a forward pass of the network, 
where the model transforms the high-dimensional input into a lower-dimensional 
representation in the target function space.

To ensure accurate coefficient estimation, we normalize the wavefunction data such that its squared norm integrates to unity over the defined grid. 
The estimated coefficients are validated by reconstructing the wavefunction 
from the predicted parameters and comparing it to the original data using 
the MSE defined in Eq. (\ref{eq:loss}). 
Typical values of the MSE are less than the order of $10^{-4}$-$10^{-3}$.
This ensures the validity of the coefficient estimation.

\subsection{Determination of the initial estimation}
By following the above procedure, a set of initial orbital candidates for Wannier functions can be obtained.
We adopt the candidate with the largest absolute coefficient $|c^{(n)}_{lm}|$ 
as the initial orbital on a single band.
For multiple bands, we determine the projection using the above method, starting from the lower-energy band.
However, if the position and the orbital overlap with the projection of the previous bands, 
we switch to the candidate with the next largest absolute coefficient.

\subsection{Code implementation}
The above procedure has been implemented in the Python code, SEAP~\cite{seap_hp}.
SEAP can be used as a post-processing tool for the output of pp.x in Quantum ESPRESSO~\cite{QE,QE_HP}.
We use \texttt{doped} package for generating interstitial sites based on the Voronoi tessellation method~\cite{Kavanagh2024} and \texttt{scipy} package for the $k$d-tree algorithm~\cite{2020SciPy-NMeth}.
For constructing and training the neural network, we employ \texttt{PyTorch}~\cite{paszke2019pytorchimperativestylehighperformance}.
The code is distributed under the MIT License and is available on GitHub. %Add ref ~\cite{HIAP}.

\section{Applications}
\label{sec:applications}
To demonstrate the usefulness of our method, we apply it to typical compounds, including a correlated metal SrVO$_{3}$, a semiconductor Si, an iron-based superconductor FeSe, a sodalite Na$_8$Al$_6$Si$_6$O$_{24}$, and an organic compound (TMTTF)$_{2}$PF$_{6}$. 
In the following, we show the results of the Wannier functions obtained by our method and compare them with those obtained by the SCDM method~\cite{Damle_MMS2018, Damle_JCC2015}. 
The pseudopotentials used in the calculations are ultrasoft pseudopotentials with the Perdew-Burke-Ernzerhof (PBE) exchange-correlation functional~\cite{PBE_1996} distributed as pslibrary~\cite{pslibrary,pslibrary_hp}.
We use Quantum ESPRESSO~\cite{QE,QE_HP} for the DFT calculations and Wannier90~\cite{Mostofi_CPC2008,Mostofi_CPC2014,Pizzi_JPCM2020,wan90_HP} for the Wannierization. 
All the Wannierizations are performed using the maximally localized Wannier functions (MLWF) procedure with 500 iteration steps~\cite{Marzari_RMP2012,Marzari_PRB1997,Souza_PRB2001}.
Crystal structures of the inorganic solids in CIF format are taken from the Materials Project database~\cite{Jain2013}.
Using cif2qewan~\cite{cif2qewan_hp}, we generate input files for Quantum ESPRESSO and Wannier90 from CIF files.
For (TMTTF)$_{2}$PF$_{6}$, we used the Quantum ESPRESSO input files corresponding to the 200 K structure, which were downloaded from the ISSP data repository\cite{tm-salts}.
Note that unless otherwise noted, we set the top of the frozen window for the Wannierization to be 2.0 eV above the Fermi energy.

\subsection{{SrVO$_{3}$}}
SrVO$_{3}$ is a correlated metal with a perovskite structure\cite{CHAMBERLAND1971243,ONODA1991281}. 
It has been extensively studied as a typical example of correlated electron materials because the band structure is quite simple: three t$_{\rm 2g}$ bands are located around the Fermi energy, and the other bands are well separated from the t$_{\rm 2g}$ bands\cite{Lechermann2006,Solovyev2006,Aryasetiawan2006,Vaugier2012,Qiao_MRWF}. 
Thus, when we perform the Wannierization, we exclude the other bands and focus on the t$_{\rm 2g}$ bands as the target bands. 

\begin{figure}
         \centering
         \resizebox*{8cm}{!}{\includegraphics{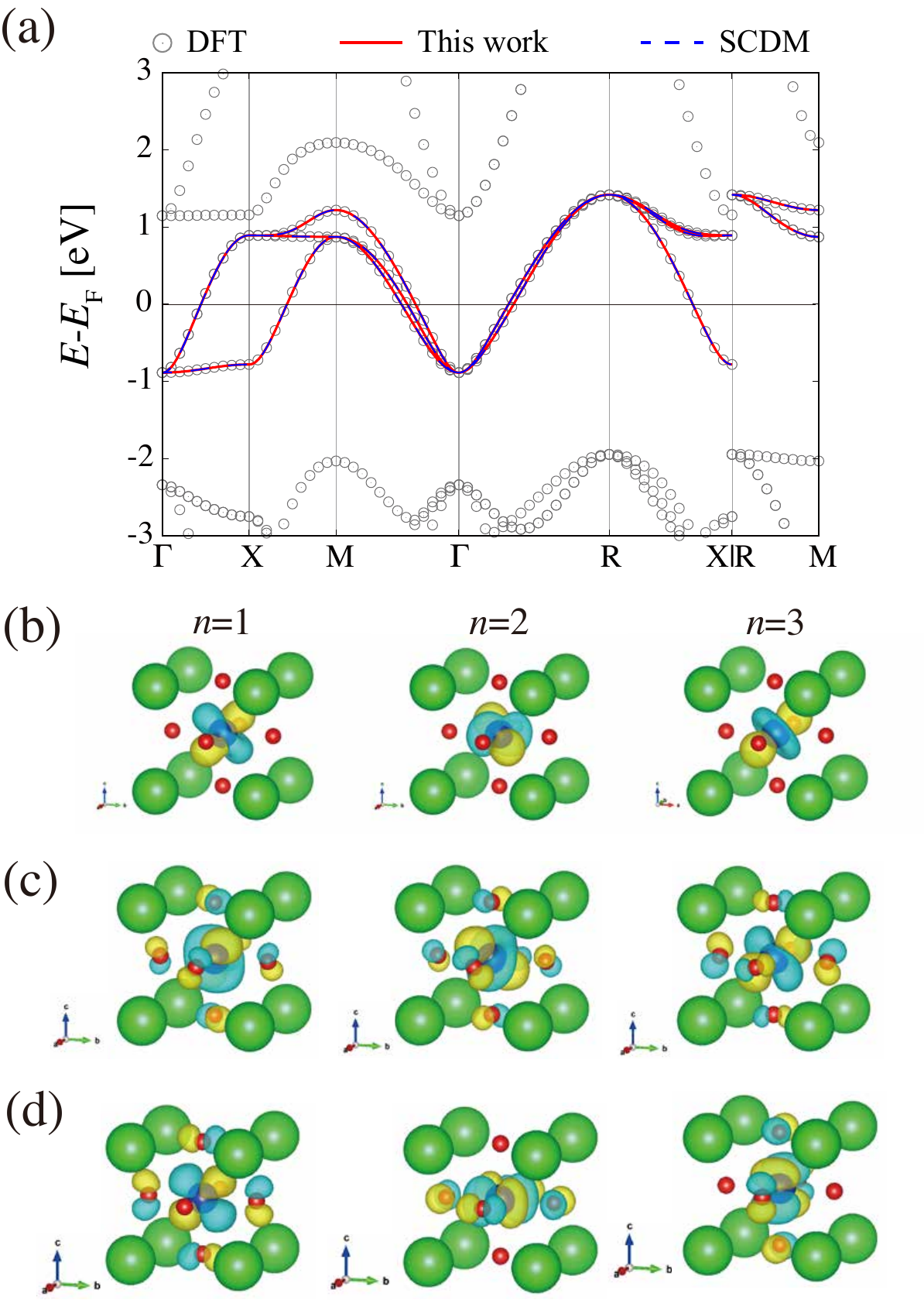}}
         \caption{(a) Band structure of SrVO$_{3}$ with Wannier interpolated bands. Circles are obtained by DFT calculations. The red line and blue dashed line are obtained by Wannier interpolation using our method and the SCDM method, respectively. The black thin line denotes the Fermi energy level.
         (b) Bloch functions at $\Gamma$ point. Green, blue and red spheres are Sr atoms, V atoms and O atoms, respectively. The yellow (sky blue) region indicates the isosurface of the positive (negative) intensity of the Bloch functions. (c-d) Wannier functions obtained by the SCDM method (c) and our method (d). Yellow and sky-blue regions mean the isosurface of the positive and negative intensity of the Wannier functions, respectively. Other notations are the same as those in panel (b).} \label{fig:svo}
\end{figure}

Figure \ref{fig:svo} (a) shows the electronic structure of SrVO$_{3}$ with Wannier interpolated bands. 
We can see that the bands are reproduced by both our method and the SCDM method, which indicates that both methods can construct the Wannier functions that well describe the target bands {\it without any prior knowledge} of them.
We also find that the spread of the Wannier functions is almost the same for both methods: the total spread $\Omega_{\rm tot}$ is about 5.682 \AA$^{2}$ for our method and 5.705 \AA$^{2}$ for the SCDM method. 
The gauge invariant part of the spread $\Omega_{\rm I}$ is about 5.682 \AA$^{2}$ for both methods.
These results suggest that the Wannier functions obtained by our method are connected to those obtained by the SCDM method under a unitary transformation and the difference between them is mainly a choice of the gauge arising from the different initial conditions. 

To further gain an insight into this point, we present real-space plotting of wave functions in 
Fig. \ref{fig:svo} (b-d). 
Figure \ref{fig:svo} (b) shows the Bloch wavefunctions at the $\Gamma$ point, which indicates that the Bloch wavefunctions are located at the V atom and have $d$-like characters. 
We find that in these calculations, $d$-like orbitals are not directed to O atoms, but to Sr atoms.
Because the SCDM method directly uses the information of the density matrix constructed by the Bloch wavefunctions as the initial guesses, the Wannier functions obtained by the SCDM method shown in Fig. \ref{fig:svo} (c) consist of $d$-like orbitals with similar characters to the Bloch wavefunctions and six oxygen $2p$ orbitals.
On the other hand, the Wannier functions obtained by our method shown in Fig. \ref{fig:svo} (d) are reflected in the hydrogenic initial guess predicted by the neural network model and thus more human-interpretable: the Wannier functions describing the low-energy degrees of freedom are vanadium $t_{2g}$ orbitals associated with four oxygen $2p$ orbitals.

For understanding the reason why our method works well, we explain the details of estimating the initial guess of the Wannier functions as follows.
First, as we mentioned in the Method section, we need to perform the clustering to determine the centers of the projected orbitals $\tilde{\bm{r}}_\nu^{(n)}$.
By performing the clustering with $d_{\rm th} = 1.5$ \AA \ in a set of the Bloch wavefunctions shown in Fig. \ref{fig:svo} (b), we find that atomic clusters are formed by each atom and all the Bloch wavefunctions are correctly assigned to the clusters. 
Because the intensity of the Bloch wavefunction in the clusters including Sr or O atom is quite weak compared with that in V's cluster, it is not considered as a candidate for the initial projection anymore. 
This result leads to the recommendation that the centers of the projected orbitals $\tilde{\bm{r}}_\nu^{(n)}$ should be located in V's cluster.
Next, we show the results of the spherical harmonics expansion of the Bloch wavefunctions.
The coefficients $c^{(n)}_{lm}$, defined in Eq.(\ref{eq:S}), of the cluster with V atom and the MSE $L$ are summarized in Table \ref{tab:svo}. 
We find that the order of the MSE $L$ is $10^{-4}$, which indicates that the neural network model can predict the coefficients of the spherical harmonics $c^{(n)}_{lm}$ well. 
We also find that the absolute values of the coefficients $|c^{(n)}_{lm}|$ of $d_{xy}$, $d_{yz}$, and $d_{xz}$ orbitals are larger than those of the others. 
This indicates that, as we expected, the $d_{xy}$, $d_{yz}$, and $d_{xz}$ orbitals are the main components of three Bloch wavefunctions and are successfully recommended as the initial projected states of the Wannier functions. 
      
\begin{table}
\caption{Coefficients of the spherical harmonics $c_{lm}^{(n)}$ for SrVO$_{3}$.
The coefficients are shown for three Bloch wavefunctions at the $\Gamma$ point ($n=1,2,3$) shown in Fig. \ref{fig:svo}(b).
The first column means the kind of orbitals with $l$ and $m$. The second to fourth columns denote the coefficients for $n=1$, $2$, and $3$.
The last row shows the mean square error (MSE) between the Bloch wavefunctions and predicted wavefunctions by our neural network model.
}
\label{tab:svo}
\centering
\begin{tabular}{c|c|c|c}
\hline
kind of orbitals ($l$ and $m$) & $n=1$ & $n=2$ & $n=3$ \\  
\hline
\hline
$s$ ($l=0,m=0$) & 0.105 & 0.065 & -0.097 \\
\hline
$p_{y}$ ($l=1,m=-1$) & 0.054 & 0.100 & 0.066 \\
$p_{z}$ ($l=1,m=0$) & 0.102 & -0.029 & 0.038 \\
$p_{x}$ ($l=1,m=1$) & -0.008 & 0.063 & 0.093 \\
\hline
$d_{xy}$ ($l=2,m=-2$) & 0.491 & -2.037 & -1.495 \\
$d_{yz}$ ($l=2,m=-1$) & -1.676 & 0.935 & -1.510 \\
$d_{z^{2}}$ ($l=2,m=0$) & -0.054 & -0.008 & -0.062 \\
$d_{xz}$ ($l=2,m=1$) & 1.517 & 1.253 & -1.544 \\
$d_{x^{2}-y^{2}}$ ($l=2,m=2$) & -0.097 & -0.083 & 0.035 \\
\hline
$f_{y(3x^2-y^2)}$ ($l=3,m=-3$) & -0.136 & -0.091 & 0.072 \\
$f_{xyz}$ ($l=3,m=-2$) & -0.003 & -0.096 & 0.118 \\
$f_{yz^2}$ ($l=3,m=-1$) & 0.007 & -0.110 & 0.125 \\
$f_{z^3}$ ($l=3,m=0$) & 0.061 & -0.050 & 0.001 \\
$f_{xz^2}$ ($l=3,m=1$) & -0.005 & -0.049 & -0.168 \\
$f_{z(x^2-y^2)}$ ($l=3,m=2$) & -0.022 & 0.014 & 0.013 \\
$f_{x(x^2-3y^2)}$ ($l=3,m=3$) & -0.089 & 0.107 & 0.060 \\
\hline
\hline
MSE & 0.0003 & 0.0003 & 0.0003 \\
\hline
\end{tabular}
\end{table}

\subsection{FeSe}
FeSe is an iron-based superconductor with the superconducting transition temperature $T_{\rm c} \sim 10$ K\cite{Hsu2008, mizuguchi2008superconductivity, hosono2015iron}. 
In previous studies, it was found that five $d$ orbitals of two Fe atoms in the unit cell are the main components of the Bloch wavefunctions around the Fermi energy\cite{Subedi2008,Lee2008,miyake2010comparison}.
As is the case with SrVO$_{3}$, the band structure around the Fermi energy is energetically isolated from the other bands.
However, the Wannierization process for FeSe is more complicated than that for SrVO$_{3}$: the band energy at the $\Gamma$ point is not degenerate and the Bloch wavefunctions are not localized only at one Fe atom, but are distributed over the unit cell.
The latter point makes it difficult to automatically determine the centers of the projected orbitals $\tilde{\bm{r}}_\nu^{(n)}$, which is resolved by the clustering method as we mentioned in Sec. \ref{sec:method-postprocess}.

\begin{figure}
         \centering
         \resizebox*{8cm}{!}{\includegraphics{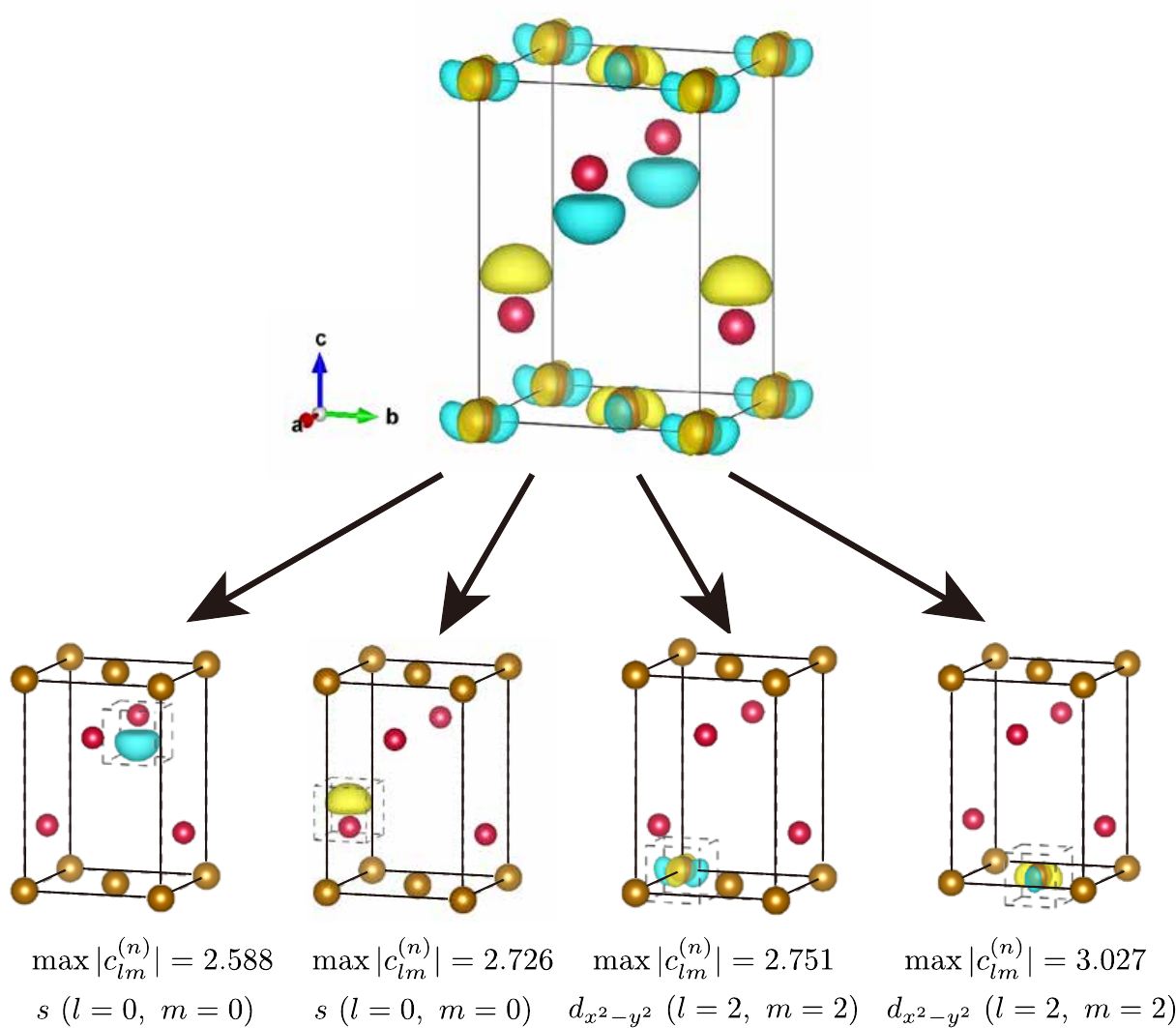}}
         \caption{Clustering of a Bloch wavefunction of FeSe around the Fermi energy.  
         Brown and deep red spheres are Fe atoms and Se atoms, respectively.
         The Bloch wavefunction (top) is clustered into four box clusters with dashed lines (bottom).
         The center of each cluster corresponds to $\tilde{\bm{r}}_\nu^{(n)}$.
         Bottom panels on the left side show components of the Bloch wavefunction within each cluster including one Se atom, and on the right side show around each cluster including one Fe atom.
         The maximum value of $|c_{lm}^{(n)}|$ of each cluster and its corresponding $l$ and $m$ are shown.} \label{fig:fese-clustering}
\end{figure}

Figure \ref{fig:fese-clustering} shows an example of the clustering process for a Bloch wavefunction of FeSe. 
The Bloch wavefunction plotted here is the eigenstate of the fourth band from the top of the isolated bands. (The band structure of FeSe will be shown later.)
We find that although the Bloch wavefunction is distributed over all the atoms in the unit cell, the clustering method successfully divides the Bloch wavefunction into four clusters surrounding each atom.
The result of estimation using the neural networks indicates that the most dominant component of each cluster is $d_{x^2-y^2}$-like orbital at the center of Fe atom or $s$-like orbital near Se atom.
Because the maximum values of $|c_{lm}^{(n)}|$ of the clusters including Fe atoms are larger than those of the clusters including Se atoms, the recommendation system concludes that appropriate initial guesses of the Wannier functions are $d_{x^2-y^2}$-like orbital at the center of either Fe atoms.
By performing the clustering for all the Bloch wavefunctions around the Fermi energy, we find that the centers of the projected orbitals $\tilde{\bm{r}}_\nu^{(n)}$ are located at the Fe atoms and the set of $d$ orbitals is recommended as the initial guesses of the Wannier functions.

\begin{figure}
   \centering
   \resizebox*{8cm}{!}{\includegraphics{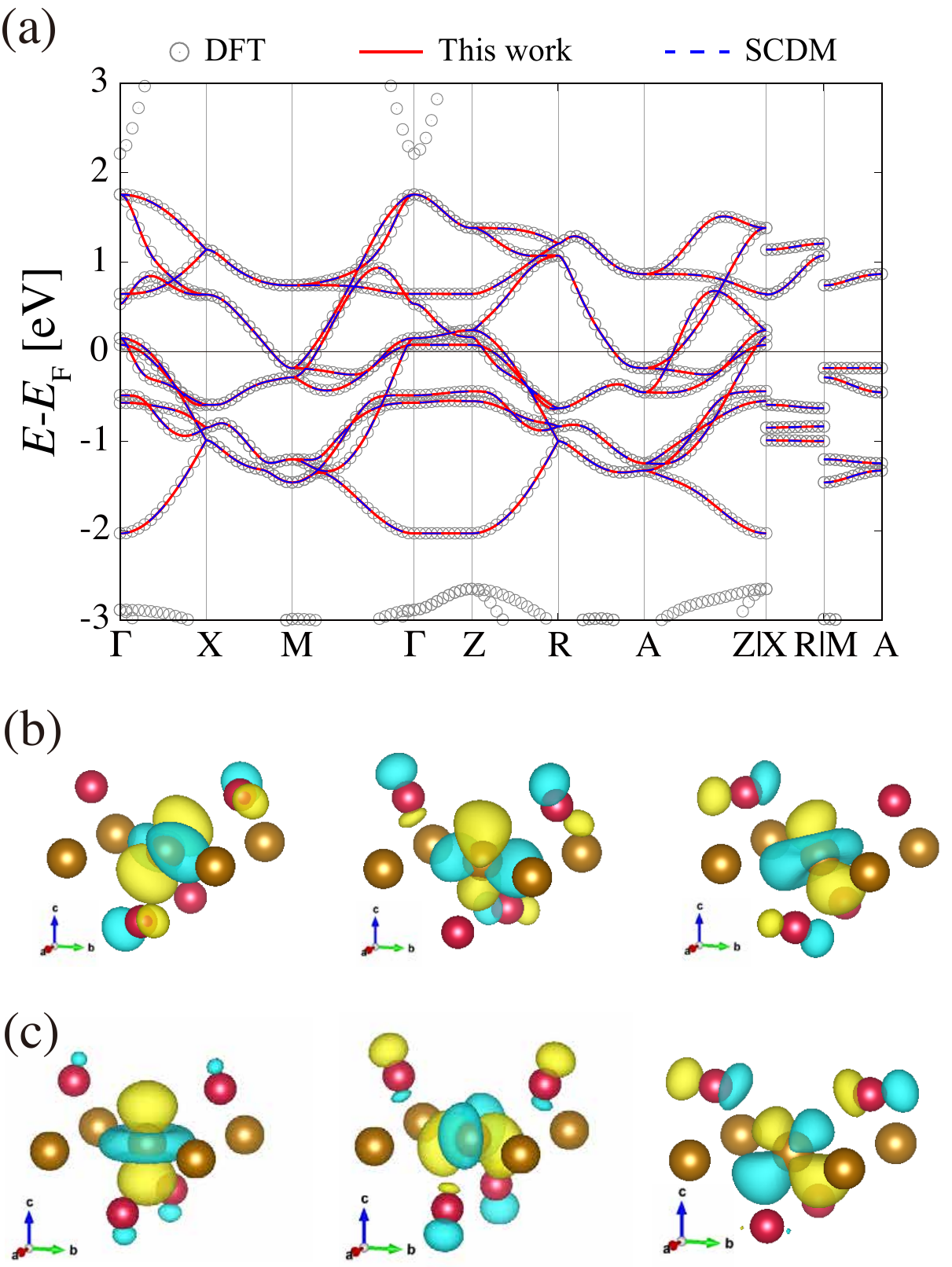}}
   \caption{(a) Band structure of FeSe with Wannier interpolated bands. 
   (b-c) Wannier functions obtained by the SCDM method (b) and our method (c).
   Brown and deep red spheres are Fe atoms and Se atoms, respectively.
   The other notations are the same as in Fig. \ref{fig:svo}.} \label{fig:fese}
\end{figure}

Figure \ref{fig:fese} (a) shows the band structure of FeSe with Wannier interpolated bands.
We find that even for the more complicated compound than SrVO$_{3}$, the bands are automatically reproduced by both our method and the SCDM method. 
The qualities of the Wannier functions are also similar to those of SrVO$_{3}$: $\Omega_{\rm tot}$ is almost the same for both methods and the gauge of the Wannier functions is more human-interpretable than those obtained by the SCDM method shown in Figs. \ref{fig:fese} (b) and (c).
We also check that the MSE between the original Bloch wavefunctions and reconstructed ones is sufficiently small, i.e., $L < 10^{-3}$, which suggests that our neural network model robustly works well for the estimation of $c_{lm}^{(n)}$ from the data set of the Bloch wavefunctions.

\subsection{Si}
Si is an insulator with a diamond structure.
It is well known that the valence and lower conduction bands are described by the $sp^{3}$ orbitals of Si atoms~\cite{Martin_2004, Kohn1973}. 
Unlike the previous two examples, the lower conduction bands are entangled with the other conduction bands, which makes it challenging to construct the Wannier functions.

The disentanglement procedure plays an important role in the Wannierization process for such entangled bands.
A widely used disentanglement procedure is to minimize the gauge-invariant spread $\Omega_{\rm I}$ of the Wannier functions on the subspace whose dimensions are less than the number of bands in the specified energy window,  which was proposed by Souza, Marzari, and Vanderbilt~\cite{Souza_PRB2001}. 
In this paper, we refer to this method as the SMV method.

Although the SMV method can be combined with the SCDM method, effects of the disentanglement can also be taken into account in the SCDM method itself when constructing ``quasi-density matrix" $P_{\bm{k}}$~\cite{Damle_MMS2018, Damle_JCC2015, vitale2020automated}.
Here, $P_{\bm{k}}$ is defined as
\begin{align}
        P_{\bm{k}} &= \sum_{n} \ket{\psi_{n\bm{k}}} f(\varepsilon_{n\bm{k}}) \bra{\psi_{n\bm{k}}},
\end{align}
where $\ket{\psi_{n\bm{k}}}$ is the Bloch wavefunction.  
$f(\varepsilon_{n\bm{k}})$ denotes the complementary error function, which is described as
\begin{align}
        f(\varepsilon_{n\bm{k}}) &= \frac{1}{2} \text{erfc}\left( \frac{\varepsilon_{n\bm{k}} - \mu}{\sigma} \right).\label{eq:erf}
\end{align}
For the case of Si, we set $\mu = -3.991$ eV and $\sigma = 5.548$ eV. 
These values are determined from fitted values $\mu_{\rm Fit}$ and $\sigma_{\rm Fit}$ in Eq. (\ref{eq:erf}) to the projectability of the states as the function of the band energy, and actual $\mu$ and $\sigma$ are obtained by $\mu=\mu_{\rm Fit}-3\sigma_{\rm Fit}$ and $\sigma = \sigma_{\rm Fit}$, as recommended in the previous study~\cite{vitale2020automated}.
We comment that in the following results, when fitting the projectability, we use only $\bm{k}$ points along the high-symmetry line instead of the sampling points in the whole Brillouin zone because the total spread of the Wannier functions is slightly better than that for the latter case.

\begin{figure}
     \centering
     \resizebox*{8cm}{!}{\includegraphics{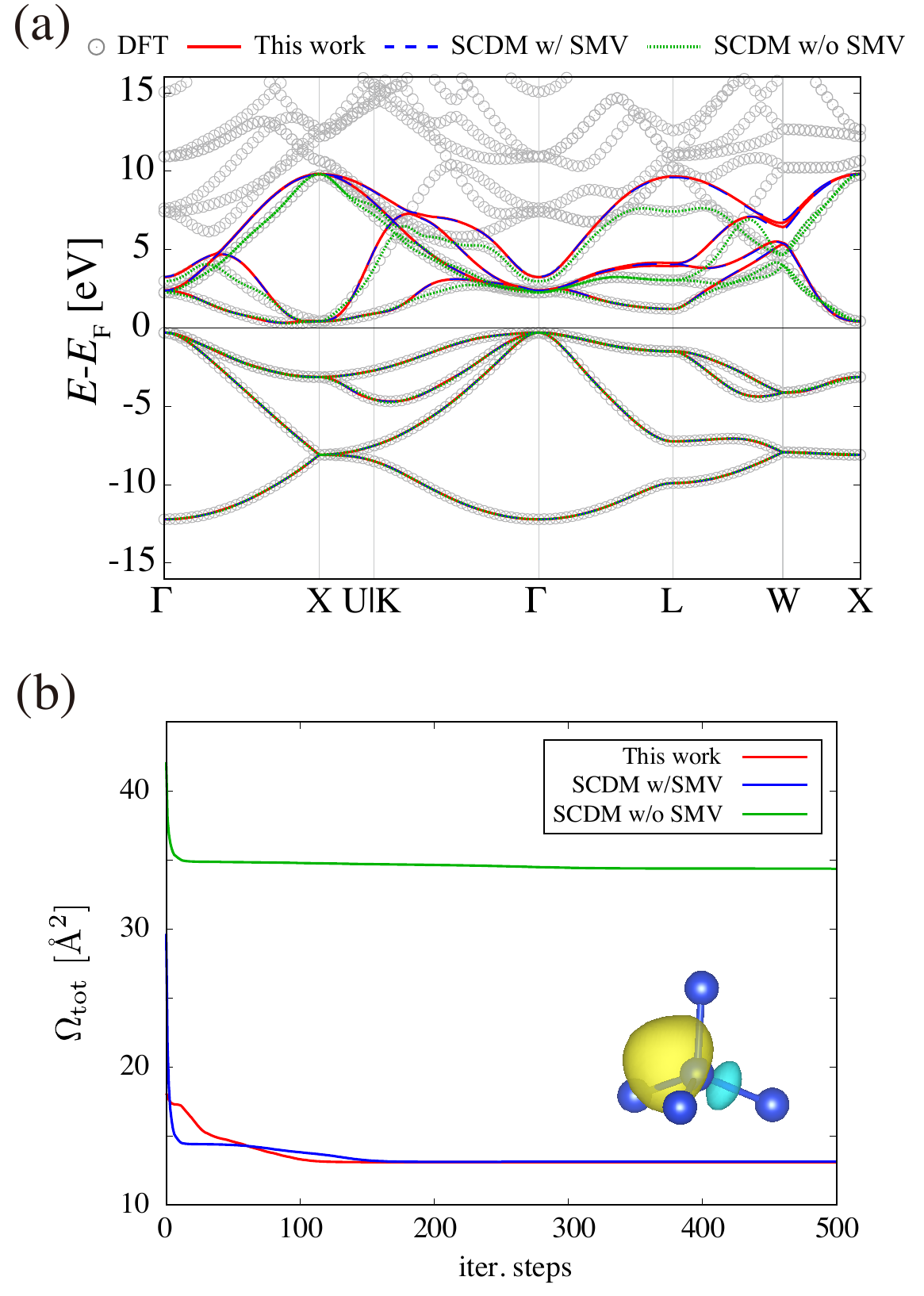}}
     \caption{(a) Band structure of Si with Wannier interpolated bands. 
     Green dotted line is obtained by the SCDM method without the SMV disentanglement procedure.
     The other notations are the same as in Fig. \ref{fig:svo}.
     (b) Total spread $\Omega_{\rm tot}$ during the MLWF procedure.
     Red, blue and green lines are obtained by our method, the SCDM method with and without the SMV disentanglement procedure, respectively. The inset represents a Wannier function obtained by our method. Blue spheres are Si atoms.
     } \label{fig:si}
\end{figure}

Figures \ref{fig:si} (a) and (b) show the band structure of Si with Wannier interpolated bands and the total spread $\Omega_{\rm tot}$ during the MLWF procedure, respectively.
We find that the valence bands are well reproduced by all the methods, while the conduction bands are not well-reproduced except for the SCDM method without the SMV disentanglement procedure.
Although the SCDM method without the SMV procedure reproduces the original band structure, the total spread $\Omega_{\rm tot}$ is much larger than that of the other methods using the SMV method. 
We also find that, regardless of the initial states, $\Omega_{\rm tot}$ using the SMV method converges to almost the same value.
This indicates that the SMV disentanglement procedure is suitable for obtaining localized Wannier functions.

We want to emphasize that our neural network model can automatically predict $s$ and three $p$ orbitals near Si atoms as recommended initial guesses.
Compared with the SCDM initialization with the SMV method, the reproducibility of the bands and $\Omega_{\rm tot}$ is of the same quality for both methods, but $\Omega_{\rm tot}$ by our method quickly converges to the minimum value. 
This is an advantage of the approach with the hydrogenic projection automatically recommended by the neural network model.
We also checked that the Wannier functions obtained by our method have $sp^3$-like characters, which is shown in the inset of Fig. \ref{fig:si} (b) as an example.

\subsection{Na$_8$Al$_6$Si$_6$O$_{24}$}
Na$_8$Al$_6$Si$_6$O$_{24}$ is the cage frame of sodium-sodalites. 
Previous studies find that the isolated bands near the Fermi energy are composed of interstitial $s$ orbitals and thus the best initial guess of the Wannier functions is $s$ orbitals at the center of the cage \cite{nakamura2009ab,Kanno_2021JPCL}.

\begin{figure}
         \centering
         \resizebox*{8cm}{!}{\includegraphics{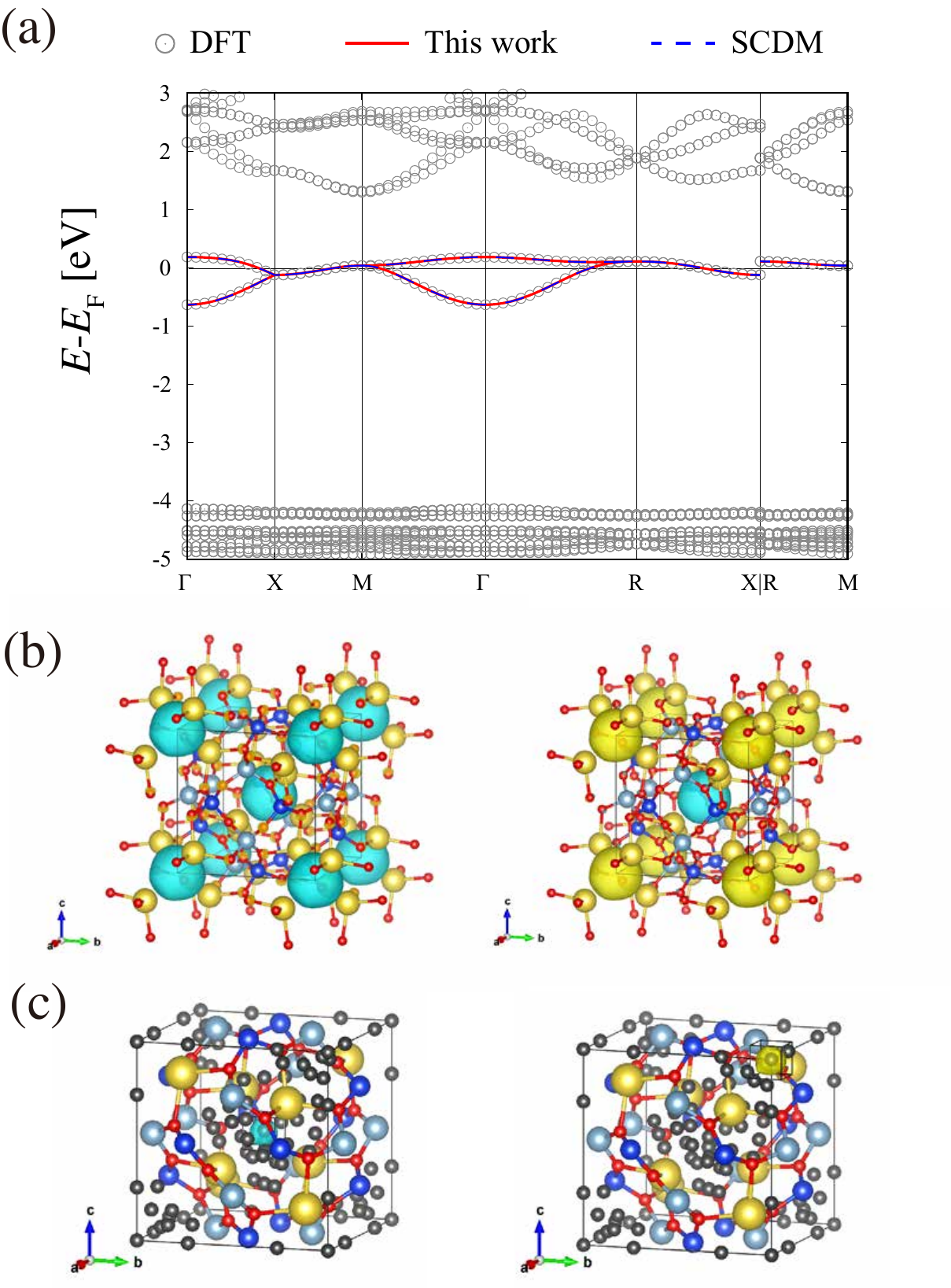}}
         \caption{(a) Band structure of Na$_8$Al$_6$Si$_6$O$_{24}$ with Wannier interpolated bands. 
         (b) Real-space plotting of the Bloch wavefunctions at the $\Gamma$ point.
         Yellow, cyan, blue, and red spheres are Na, Al, Si, and O atoms, respectively. 
         (c) Clustering of a Bloch wavefunction.
         Black spheres represent interstitial sites generated by the Voronoi tessellation method with $d_{{\rm th}}^{{\rm ai}} = 2.0 {\rm \AA}$.
         The other notations are the same as in Fig. \ref{fig:svo}.
         } \label{fig:sodalite}
 \end{figure}

Figures \ref{fig:sodalite} (a) and (b) show the band structure of Na$_8$Al$_6$Si$_6$O$_{24}$ near the Fermi energy and the real-space plotting of the Bloch wavefunctions at the $\Gamma$ point, respectively.
We find that the Bloch wavefunctions of the low-energy bands are located at the center of the cage and have $s$-like characters, which is consistent with the previous studies\cite{nakamura2009ab,Kanno_2021JPCL}. 
The results also indicate that the bands near the Fermi energy are formed by the bonding/antibonding states of the interstitial $s$ orbitals on the body-centered cubic lattice.
The result of the clustering process is shown in Fig. \ref{fig:sodalite} (c).
We find that the clustering process successfully divides the Bloch wavefunction into two clusters, which are located at the center of the cage.
We confirm that the neural network model predicts that the most dominant component of each cluster is $s$-like orbital and $\tilde{\bm{r}}_\nu^{(n)}$ becomes near the center of the cage.

\begin{figure}
   \centering
   \resizebox*{8cm}{!}{\includegraphics{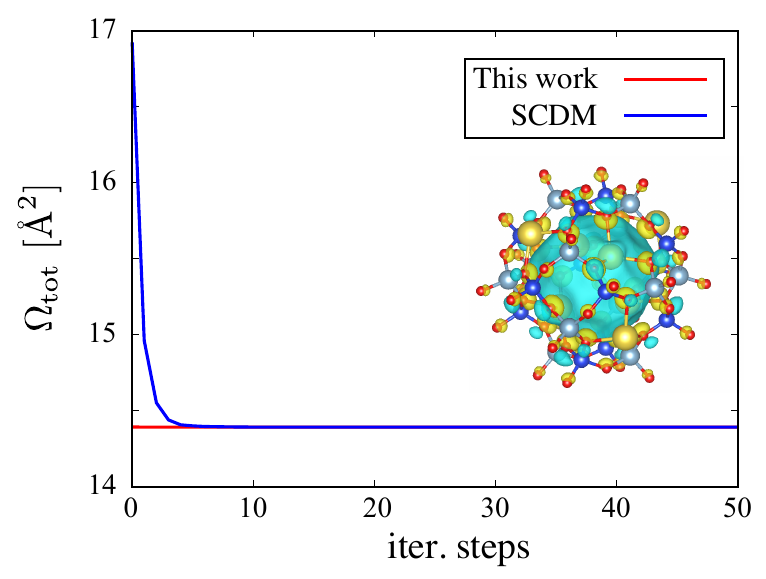}}
   \caption{Total spread $\Omega_{\rm tot}$ during the MLWF procedure for Na$_8$Al$_6$Si$_6$O$_{24}$.
   The red and blue lines are obtained by our method, the SCDM method, respectively.
   The inset represents a Wannier function obtained by our method.} \label{fig:sodalite-spread}
\end{figure}

Figure \ref{fig:sodalite-spread} shows the total spread $\Omega_{\rm tot}$ during the MLWF procedure.
Although the total spread $\Omega_{\rm tot}$ using the SCDM method quickly converges to the minimum value, $\Omega_{\rm tot}$ using the hydrogenic projection has already been well localized before the MLWF procedure starts.
This result indicates that our method can successfully propose appropriate initial guesses of the Wannier functions even for nanostructured compounds with interstitial sites like Na$_8$Al$_6$Si$_6$O$_{24}$.

\subsection{(TMTTF)$_2$PF$_6$}
Finally, we show the application to a different class of compounds, i.e., molecular solids.
In molecular solids, van der Waals interactions between molecules give rise to periodic arrangements, analogous to those observed in inorganic crystals.
Although many elements are included in the unit cell of molecular solids and thus one may expect that the band structures are complicated compared with those of inorganic solids, in most cases, the number of bands across the Fermi energy is only a few and the band structure around the Fermi energy is quite simple~\cite{seo2004}.
This is because molecular orbitals govern the energy hierarchy of electronic structures on behalf of atomic orbitals.
Thus, it is difficult to automatically select appropriate initial guesses of the Wannier functions from the hydrogenic projections. 

TMTTMF salts are widely recognized as prototypical strongly correlated molecular conductors, in which the delicate balance among charge, spin, and lattice degrees of freedom gives rise to a variety of phases~\cite{Jerome01012002}. 
Their electronic properties exhibit strong sensitivity to external parameters such as pressure~\cite{PhysRevResearch.6.043308} and anion substitution~\cite{PhysRevLett.108.096402, PhysRevLett.131.036401}, leading to rich phase diagrams with charge-ordered, spin-ordered, and superconducting states. Owing to this tunability, (TMTTF)$_2$PF$_6$ and related compounds have been extensively studied both experimentally and theoretically as model systems of strongly correlated low-dimensional materials.

\begin{figure}
   \centering
   \resizebox*{8cm}{!}{\includegraphics{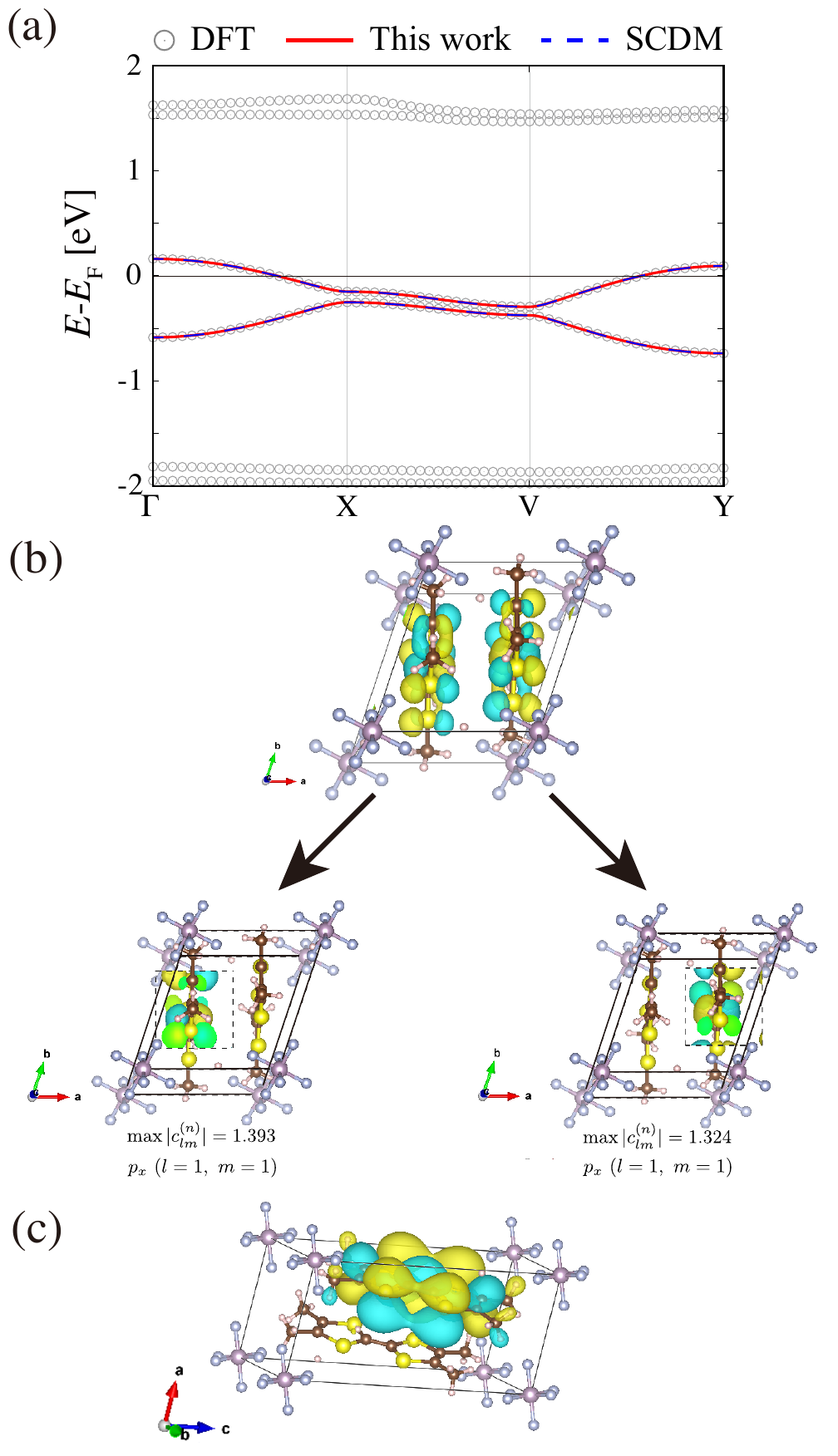}}
   \caption{(a) Band structure of (TMTTF)$_{2}$PF$_{6}$ with Wannier interpolated bands. Notations are the same as in Fig. \ref{fig:svo} (a).
   (b) Clustering of a Bloch wavefunction of (TMTTF)$_{2}$PF$_{6}$ around the Fermi energy. 
   Moleculars on the corner of the unit cell are PF$_{6}$ molecules and the other ones are TMTTF molecules.
   The other notations are the same as in Fig. \ref{fig:fese-clustering}.
   (c) A Wannier function obtained by our method.
   } \label{fig:tmttf}
\end{figure}

Figure \ref{fig:tmttf} (a) shows the band structure of (TMTTF)$_{2}$PF$_{6}$ with Wannier interpolated bands. 
We can clearly see in Fig. \ref{fig:tmttf} (a) that two bands around the Fermi energy are well isolated from the other bands and we set them as the target bands of the Wannierization.
Therefore, the main tasks of the Wannierization are to select a kind of hydrogenic orbitals appropriately and to determine the centers of the projected orbitals $\tilde{\bm{r}}_\nu^{(n)}$. 
We find that, even for the molecular solid, the Wannier interpolated bands using both our method and the SCDM method are well consistent with the original band structure, which suggests that both methods can appropriately select initial guesses of the MLWF procedure.

Figure \ref{fig:tmttf} (b) shows the clustering process of one of the Bloch wavefunctions around the Fermi energy. 
The clustering method with $d_{\rm th} = 2.0$ \AA \ successfully divides the Bloch wavefunction into three clusters, which are located around two TMTTF and PF$_{6}$ molecules, respectively.
Since the intensity of the Bloch wavefunction is sufficiently small at the PF$_{6}$ molecule, we discard the estimation process of the coefficients $c_{lm}^{(n)}$ for the PF$_{6}$ molecule and can focus only on the TMTTF molecule. 
Our neural network model predicts that the most dominant component of each cluster is $p_{x}$ orbital and $\tilde{\bm{r}}_\nu^{(n)}$ becomes near the center of TMTTF molecule.

The Wannier function obtained by our method is shown in Fig. \ref{fig:tmttf} (c) as an example. 
The Wannier function is successfully localized at each TMTTF molecule and is consistent with the previous study~{\cite{PhysRevLett.131.036401}}. 
We verified that the quality of the Wannier function is nearly identical to that of the SCDM method.

\section{Discussion and Summary}
\label{sec:summary}
In summary, to overcome the difficulty of selecting appropriate initial guesses of the MLWF procedure based on the hydrogenic projection approach, we have proposed a method to automatically select them using a neural network model.
Our scheme consists of three steps: (i) clustering the wavefunction data to determine the centers of the orbitals $\tilde{\bm{r}}=\bm{r}-\tilde{\bm{r}}_{\nu}^{n}$, (ii) estimating the coefficients $c_{lm}^{(n)}$ of the spherical harmonics expansion using the neural network model and (iii) determining the initial guesses of the MLWF procedure based on the clustering results and the estimated coefficients $c_{lm}^{(n)}$.
By applying it to a wide variety of compounds, including not only the inorganic solids but also the molecular solids, we have shown that our recommendation system works well in all cases.

In our scheme, one needs to set the threshold distance $d_{\rm th}$ and $l_{\rm c}$ in the clustering process to obtain appropriate initial guesses for the target bands one specifies.
We do not know whether this set of parameters is suitable for other compounds or not.
The verification of the parameters for other compounds is desirable via further work, such as high-throughput calculations or symmetry-adapted extensions~\cite{Sakuma2013,Koretsune2023}, but will be left for a future study.

\begin{acknowledgments}
We acknowledge Tetsuya Shoji, Noritsugu Sakuma and Hisazumi Akai for fruitful discussions and important suggestions.
We use VESTA for visualizing the crystal structures and wavefunctions~\cite{vesta}.
A part of this work is financially supported by TOYOTA MOTOR CORPORATION.
This work has also  been supported by a Grant-in-Aid for Scientific Research
(Nos. JP22K03447, JP23H03818, JP23H04869, JP23K13055, JP25K22013, and JP25H01403) from Ministry of Education, Culture, Sports, Science and Technology, Japan.
TM is also supported by JST FOREST (JPMJFR236N).
TK is also supported by JST-Mirai Program (JPMJMI20A1), JST-ASPIRE (JPMJAP2317) and Center for Science and Innovation in Spintronics (CSIS), Tohoku University. 
\end{acknowledgments}

\bibliography{main}
\end{document}